\documentclass[%
 reprint,
 superscriptaddress,
 nolongbibliography,
showpacs,
amsmath,amssymb,
aps,
pra,
 twocolumn
]{revtex4-1}
\usepackage{graphicx}
\usepackage{dcolumn}
\usepackage{bm}
\usepackage{color}
\usepackage{verbatim}

\usepackage[usenames,dvipsnames]{xcolor}
\usepackage[breaklinks, pdftex, hyperfootnotes=true, pdfpagelabels, bookmarks, pageanchor]{hyperref}
\hypersetup{%
	colorlinks=true, linktocpage=true, pdfstartpage=1, pdfstartview=FitH, pdfborder={0 0 0},%
	breaklinks=true, pdfpagemode=UseNone, pageanchor=true, pdfpagemode=UseOutlines,%
	plainpages=false, bookmarksnumbered, bookmarksopen=true, bookmarksopenlevel=1,%
	hypertexnames=true, pdfhighlight=/O,
	urlcolor=blue, linkcolor=blue, citecolor=blue, 
}

\begin{document}

\preprint{APS/123-QED}

\title{Reduction of the energy-gap scaling by coherent catalysis in models of quantum annealing}

\author{Yang Wei Koh}
\email{patrickkyw@gmail.com}

\affiliation{Institute of Innovative Research, Tokyo Institute of Technology, Nagatsuta-cho, Midori-ku, Yokohama 226-8503, Japan}

\author{Hidetoshi Nishimori}
\affiliation{Institute of Innovative Research, Tokyo Institute of Technology, Nagatsuta-cho, Midori-ku, Yokohama 226-8503, Japan}
\affiliation{Graduate School of Information Sciences, Tohoku University, Sendai 980-8579, Japan}
\affiliation{RIKEN, Interdisciplinary Theoretical and Mathematical Sciences (iTHEMS),
Wako, Saitama 351-0198, Japan}

\date{\today}

\begin{abstract}

Non-stoquastic drivers are known to improve the performance of quantum annealing by reducing first-order phase transitions into second-order ones in several mean-field-type model systems. Nevertheless, statistical-mechanical analysis shows that some target Hamiltonians still exhibit unavoidable first-order transitions even with non-stoquastic drivers, making them difficult for quantum annealing to solve. Recently, a mechanism called coherent catalysis was proposed by Durkin [Phys. Rev. A \textbf{99}, 032315 (2019)], in which he showed the existence of a particular point on the line of first-order phase transitions where the energy gap scales polynomially as expected for a second-order transition. We show by extensive numerical computations that this phenomenon is observed in a few additional mean-field-type optimization problems where non-stoquastic drivers fail to change the order of phase transition in the thermodynamic limit. This opens up the possibility of using coherent catalysis to  search for exponential speedups in systems previously thought to be exponentially slow for quantum annealing to solve.

\end{abstract}
\maketitle


\section{Introduction}
\label{sec.I}

Quantum annealing (QA) \cite{Kadowaki1998,Brooke1999,Farhi2001,Santoro2002,Santoro2006,Das2008,Morita2008,Albash2018,Hauke2019} is a method for solving combinatorial optimization problems by making use of the tunneling effects induced by quantum fluctuations to search for the optimal solution. An important challenge facing QA is that for certain classes of problems one encounters a first-order phase transition during the annealing process. At a first-order transition, the energy gap $\Delta=E_1-E_0$ between the energies of the ground and first-excited states of the system, $E_0$ and $E_1$, generally decreases exponentially with system size, resulting in an exponentially long annealing time,  according to the adiabatic theorem of quantum mechanics \cite{Kato:50,Jansen:07,lidar:102106}. One such example is the ferromagnetic $p$-spin model ($p$-spin ferromagnet) whose target Hamiltonian is given by 
\begin{equation}
H=-N{J_z}^p
\label{eq.sec.I.01}
\end{equation}
where the power $p\ge 2$ is an integer, $N$ is the total number of spins, $J_{\alpha}=N^{-1}\sum_{i=1}^N \sigma^{\alpha}_i$, and $\sigma^{\alpha}_i$ $(\alpha=x,y,z)$ is the $\alpha$-component of the Pauli matrix for the $i$th spin. J\"{o}rg \emph{et al.} \cite{Jorg10} showed that, if QA is performed with just a transverse field $J_x$, the system undergoes a first-order phase transition for $p>2$ and is therefore hard to solve although the solution is trivially ferromagnetic.  It was later found that, by introducing non-stoquastic drivers of the form $+{J_x}^q$ ($q\ge 2$) into the annealing Hamiltonian, many of the first-order phase transitions that plagued a simple transverse field annealing in the $p$-spin ferromagnet can be avoided and be replaced by gentler second-order ones whose annealing times are expected to scale only polynomially with the system size \cite{Seki12,Seoane12,Seki2015,Nishimori17,Takada2019}. An intractable exception remains, however, for the case of $p=3$ where mean-field analysis, which is valid in the thermodynamic (large-size) limit, suggests that first-order transitions cannot be avoided even with a non-stoquastic driver \cite{Seki12,Seoane12,Nishimori17}. It is also known that the $p$-spin ferromagnet with $p\ge 4$ becomes difficult to solve even with a non-stoquastic driver in the presence of random longitudinal fields \cite{Ichikawa:Thesis:2014}.

In a recent development, Durkin \cite{Durkin19} examined the $p=3$ case in detail and found that it is in fact actually possible to reduce the computation time from exponential to polynomial as a function of the system size with the help of a non-stoquastic driver $+{J_x}^2$ if one carefully controls the system parameters as functions of the system size. By exact numerical diagonalization and mapping of the problem to an effective single-particle problem, he showed that, along the curve of minimum energy gap that separates the paramagnetic and ferromagnetic phases in the phase diagram, there exists a particular point where the gap scales polynomially with the system size. At this point on the phase transition curve, kinetic effects due to the transverse antiferromagnetic interaction (non-stoquastic driver) delocalizes the ground-state wavefunction into a form that simultaneously overlaps the two minima of the classical potential, thereby softening the first-order phase transition into an effectively second-order one. This process is termed `coherent catalysis' by the author, and opens up the possibility of searching for quantum speedups in systems which are predicted by the mean-field theory to exhibit only first-order transitions. 

In Ref. \cite{Durkin19}, the use of coherent catalysis to change the gap scaling from exponential to polynomial in QA was demonstrated exclusively for the $p=3$ model. Since this result is a consequence of intricate balance among various terms in the Hamiltonian, a natural questions arises whether or not the mechanism of coherent catalysis applies to other problems. The goal of the present paper is to answer this question. The models we have chosen to study here are difficult for QA in the sense that the conventional mean-field theory predicts that a first-order phase transition is unavoidable even in the presence of non-stoquastic drivers. Our approach is computational and we shall focus mainly on computing the energy gap by diagonalizing the Hamiltonian numerically and obtaining its scaling. Our results indicate that the mechanism of coherent catalysis works also in our problems.

The rest of the paper is organized as follows. Sections \ref{sec.II} to \ref{sec.IV} are devoted to three different models. In Sec. \ref{sec.II} we extend the Hamiltonian studied in Ref. \cite{Durkin19} to include many-body transverse interactions. This is to verify that the result reported in Ref. \cite{Durkin19} is not just peculiar to the case of $+{J_x}^q$ with $q=2$ but holds for higher powers of $q$ as well. In Sec. \ref{sec.III} we go beyond the simple 3-spin ferromagnet and study a target Hamiltonian whose energy landscape exhibits a barrier and a local minimum. In Sec. \ref{sec.IV} we study the $p$-spin ferromagnet in a random longitudinal field. A previous mean-field study of this model has shown that the random longitudinal field induces a first-order transition that is not lifted by a non-stoquastic driver even for $p\ge 4$ where the difficulty can be removed by a non-stoquastic driver in the absence of random field \cite{Ichikawa:Thesis:2014}. Section \ref{sec.V} discusses and concludes the paper.

\section{The 3-spin ferromagnet with many-body transverse interactions}
\label{sec.II}

The first model we study is the 3-spin ferromagnet with many-body transverse interactions given by the Hamiltonian
\begin{equation}
H_{\mathrm{I}}(\Gamma,\kappa)
=
-\kappa(1-\Gamma){J_z}^3
-\Gamma J_x
+ (1-\Gamma)(1-\kappa){J_x}^q
\label{eq.sec.II.01}
\end{equation}
where $-{J_z}^3$ is the target Hamiltonian, $\Gamma$ and $\kappa$ are parameters controlling the strengths of the transverse field $J_x$ and the non-stoquastic driver term $+{J_x}^q$, respectively, and the integer power $q\ge 2$ generates the $q$-body transverse interactions. We have used the above parameterization by $\kappa$ and $\Gamma$ following the convention adopted in Ref. \cite{Durkin19}. We have also chosen the Hamiltonian to be of order $N^0$ in the present section for consistency with Ref. \cite{Durkin19}.

\begin{figure}[htp]
\begin{center}
\includegraphics[scale=0.7]{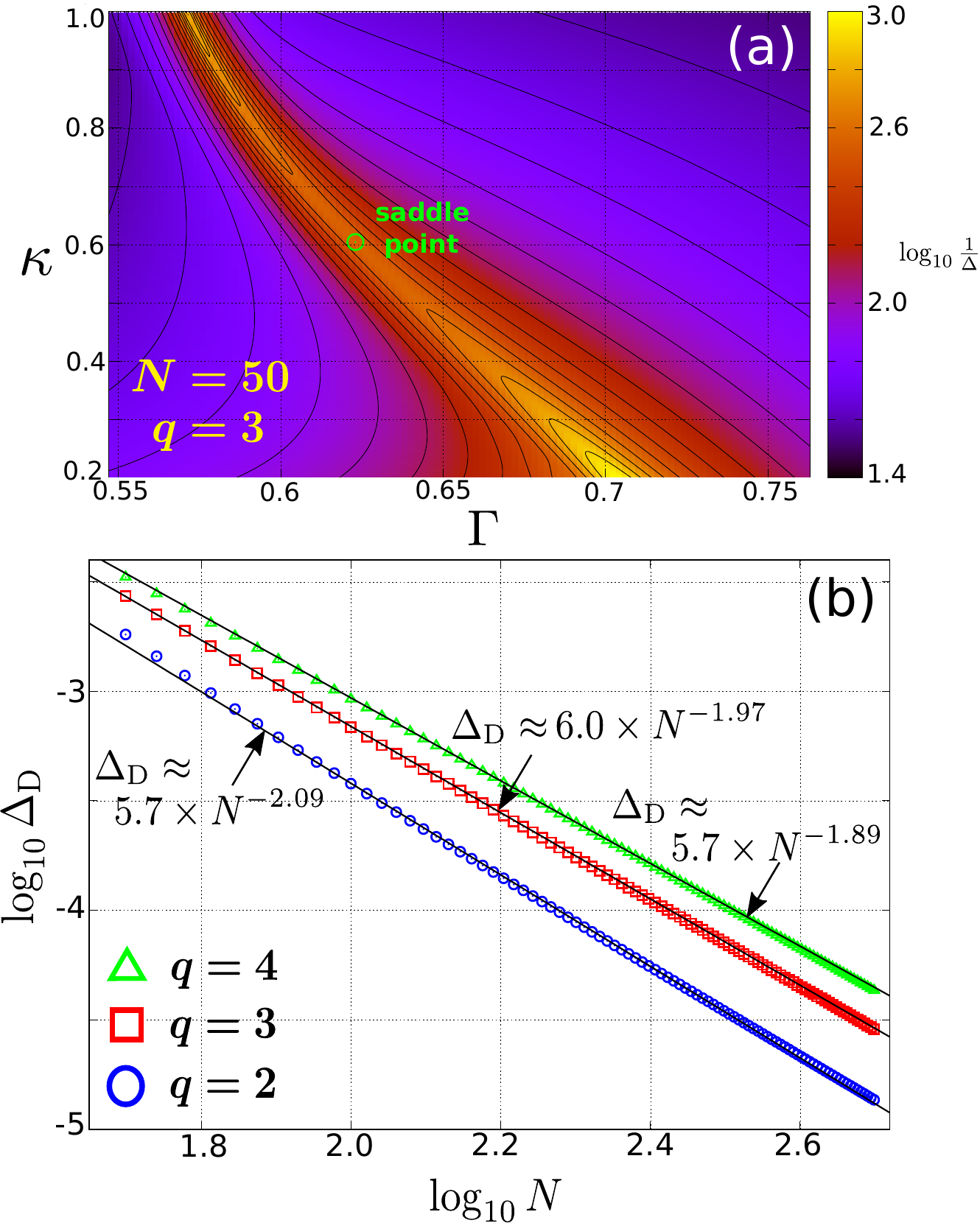}
\caption{Behavior of the energy gap of the model Eq. (\ref{eq.sec.II.01}). (a) Density plot of the logarithm of the inverse of the energy gap, $\log_{10}(1/\Delta)$, as a function of the parameters $\kappa$ and $\Gamma$ for $q=3$ and $N=50$. The location of the saddle point on this `gap landscape' is shown as a green circle. Equipotential contours of $\log_{10}(1/\Delta)$ are overlaid to aid in the visualization of the saddle point. Annealing starts from $\Gamma=1$, $\kappa$ arbitrary and ends at $\Gamma=0, \kappa=1$. (b) The energy gap evaluated at the saddle point, denoted $\Delta_{\mathrm{D}}$, decreases polynomially as the system size $N$ increases. Results for $q=2$ (studied in Ref. \cite{Durkin19}) and higher powers $q=3$ and $4$ are shown. The solid black line corresponding to each $q$ is obtained by fitting a straight line through all the data points.}
\label{fig.01}
\end{center}
\end{figure}

When $q=2$, $H_{\mathrm{I}}$ reduces to the model studied in Ref. \cite{Durkin19}. For QA, one starts from $\Gamma=1$ and $\kappa$ arbitrary and tries to find a path in the $\Gamma$-$\kappa$ parameter space to reach $(\Gamma,\kappa)=(0,1)$ where the target classical Hamiltonian is recovered. A mean-field analysis of this system shows that the $\Gamma$-$\kappa$ plane is separated into two regions by a first-order transition line \cite{Seoane12}. Although the precise location of this line depends on $q$, qualitatively, it lies within the curved orange-yellow segment of the heat map shown in Fig. \ref{fig.01}(a).

According to the  mean-field theory, during the annealing process the system must cross this line and undergo a first-order phase transition, and therefore the gap is naively expected to decrease exponentially as a function of the system size. This problem with general $q$ was touched upon in Ref. \cite{Durkin19} but was not analyzed in detail except for the case of $q=2$.  Another reason that we study this model is to confirm the accuracy of our numerical method to discern exponential and polynomial gap closings, which will turn out to be serious also in other problems to be discussed later.

The Hamiltonian $H_{\mathrm{I}}$ commutes with the total angular momentum operator ${J_x}^2+{J_y}^2+{J_z}^2$. Since quantum annealing starts from the state with only the transverse field term $\Gamma=1$ in the Hamiltonian Eq. (\ref{eq.sec.II.01}), one can simply diagonalize $H_{\mathrm{I}}$ in the sector with total angular momentum $N/2$. The result is summarized in Fig. \ref{fig.01}(a) as the density plot of the logarithm of the energy gap  as a function of $\Gamma$ and $\kappa$ for the case of $q=3$ with $N=50$. Qualitatively, the `gap landscape' here is similar to that of $q=2$ reported in Ref. \cite{Durkin19}. The narrow orange-yellow curve slanting down the middle is where the energy gap is smallest, and separates the ferromagnetic phase on the left from the paramagnetic phase on the right. In the limit $N\rightarrow\infty$, the minimum gap curve asymptotically approaches the first-order phase transition curve given by the mean-field theory \cite{Seoane12}. It therefore seems impossible to avoid a first-order transition to reach the final state at $\Gamma=0, \kappa=1$ from the initial state at $\Gamma=1$, $\kappa$ arbitrary, suggesting an exponentially long computation time.

An outstanding feature about this gap landscape is the existence of a saddle point along the minimum gap curve where the gap is a maximum as pointed out for the case $q=2$ in Ref. \cite{Durkin19}. In Fig. \ref{fig.01}(a), this point is indicated by a green circle. It was reported for $q=2$ in Ref. \cite{Durkin19} that the gap evaluated at this point obeys a different, polynomial, scaling compared to the other points with exponential scaling along the minimum gap curve. To see if the same happens in the present case of $q=3$, we tracked the saddle point as $N$ increases, which is known to scale as $\kappa_c\propto 1/\sqrt{N}$ for $\kappa\ll 1$ for $q=2$. Numerical identification of a saddle point in the very steep gap landscape is  a highly non-trivial task, and we used the Newton-Raphson algorithm to find the point $(\Gamma_{\mathrm{D}},\kappa_{\mathrm{D}})$ where the first derivatives (with respect to $\Gamma$ and $\kappa$) of the function $\log_{10}(1/\Delta)$ vanishes. The gradient and Hessian required are also computed numerically, by taking finite differences of $\log_{10}(1/\Delta)$ along the $\Gamma$ and $\kappa$ directions. As the initial condition for the Newton-Raphson algorithm, it is preferable to choose a point which is near to the solution (such as the saddle point from the previous $N$).

The results for the scaling of the gap evaluated at the saddle point, denoted as $\Delta_{\mathrm{D}}$,  are shown in Fig. \ref{fig.01}(b) for $q=2,3,$ and 4 (from $N=50$ to 500). The three solid lines shown are obtained by fitting the data points (from all $N$) to a polynomial function of the form $c_1 N^{-c_2}$ ($c_1,c_2:$ fitting parameters). It is seen that $\Delta_{\mathrm{D}}$ decreases polynomially with increase in system size for all three $q$ values. The scaling result we obtained for $q=2$ also agrees with the analytic result $3.4\times N^{-2}$ derived in Ref. \cite{Durkin19} (which holds in the large $N$ limit)
\footnote{The difference which we obtained for the value of $c_1$ could be attributable to us including data with `small' $N$ values (down to $N=50$) in our line fitting.}.
Elsewhere along the minimum gap curve, the energy gap closes exponentially fast with $N$, in accordance with the prediction of the mean field theory. This has been discussed for the case of $q=2$ in Ref. \cite{Durkin19}, and which we also verified for the cases of $q=3$ and 4 based on our numerical calculations. We have thus confirmed that the method of coherent catalysis works also in this slightly different problem and that our numerical technique succeeds to adequately follow the saddle point in the gap landscape where the valley of minimum gap has a very sharp, edge-like, shape.


\section{A target Hamiltonian exhibiting an energy barrier}
\label{sec.III}

\begin{figure}[h]
\begin{center}
\includegraphics[scale=0.7]{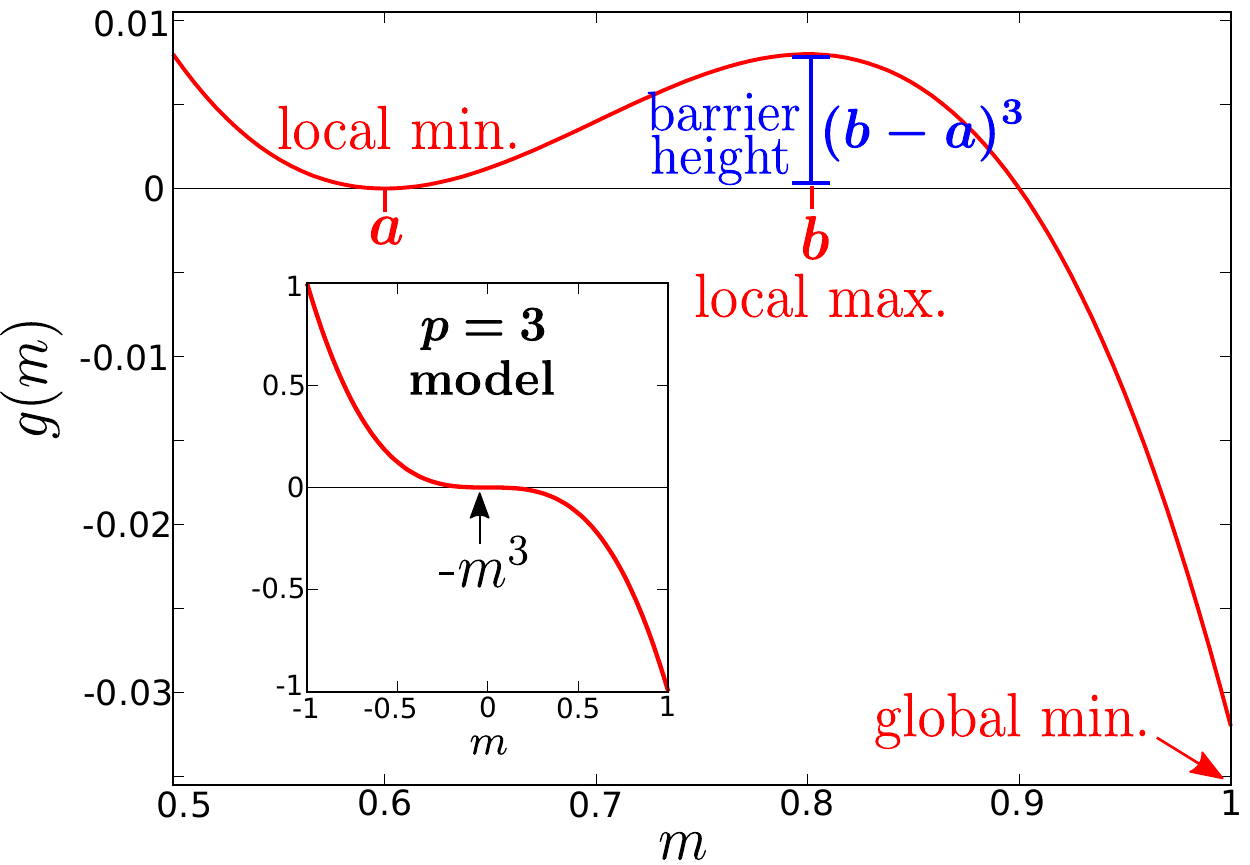}
\caption{Graph of the potential function $g(m)$ which is supposed to be minimized by the annealing Hamiltonian Eq. (\ref{eq.sec.III.01}). The parameters $a$ and $b$ in Eq. (\ref{eq.sec.III.02}) are the positions of the local minimum and maximum, respectively. The height of the barrier measured relative to the potential at the local minimum is $(b-a)^3$. Quantum annealing seeks to reach the global minimum solution located at $m=1$. The plot shows the case of $a=0.6\,,\,b=0.8$. For comparison, the inset shows the potential function of the $p=3$ model, which is just a simple $-m^3$ function and exhibits no local minima.}
\label{fig.02}
\end{center}
\end{figure}

One may ask if the very simple structure of the target Hamiltonian $-{J_z}^3$ with a single trivial minimum may be the main culprit of the success of coherent catalysis. We therefore study a model with a local minimum separated from the global minimum by an energy barrier in the target Hamiltonian. The total Hamiltonian is given by
\begin{equation}
\frac{1}{N}H_{\mathrm{II}}(s,\lambda)
=
s
\left[
\lambda\,
g\left(J_z\right)
+
(1-\lambda)
{J_x}^2
\right]
-
(1-s)J_x
\label{eq.sec.III.01}
\end{equation}
where $g(J_z)$ is the target Hamiltonian, and $s$ and $\lambda$ are parameters controlling the strengths of the transverse field $J_x$ and the antiferromagnetic interaction (non-stoquastic driver) $+{J_x}^2$, respectively. The parameterization by $s$ and $\lambda$ follows the convention of Refs. \cite{Seki12,Seoane12,Nishimori17}. The function $g(m)$ is given by
\begin{equation}
g(m)=-2(m-a)^3 - 3(a-b)(m-a)^2
\label{eq.sec.III.02}
\end{equation}
and its graph is shown in Fig. \ref{fig.02}
\footnote{The function $g(m)$ was proposed, and its QA studied in detail using mean-field theory, by Yu Yamashiro \cite{Yamashiro2020}}.
The parameters $a$ and $b$ are the coordinates of the local minimum and local maximum, respectively. The height of the energy barrier measured relative to the potential at $m=a$ is $(b-a)^3$. For comparison, the graph of the target Hamiltonian of the 3-spin ferromagnet is shown in the inset. The latter is a monotonic function with only one minimum solution at $m=1$. From an algorithmic point of view, finding the optimal minimum of $g(m)$ should be more non-trivial than that of $-m^3$ due to the presence of the local minimum.  For the QA of $H_{\mathrm{II}}$, one starts from  $s=0$, $\lambda$ arbitrary and finishes at $(\lambda,s)=(1,1)$ where the target is recovered. A mean-field analysis of this model once again predicts that the system must encounter an unavoidable first-order phase transition during the annealing process \cite{Yamashiro2020}. In the following, we examine this model from the perspective of coherent catalysis.

\subsection{Saddle point with polynomial scaling of the energy gap}
\label{sec:saddlepoint_poly}

\begin{figure}[h]
\begin{center}
\includegraphics[scale=0.7]{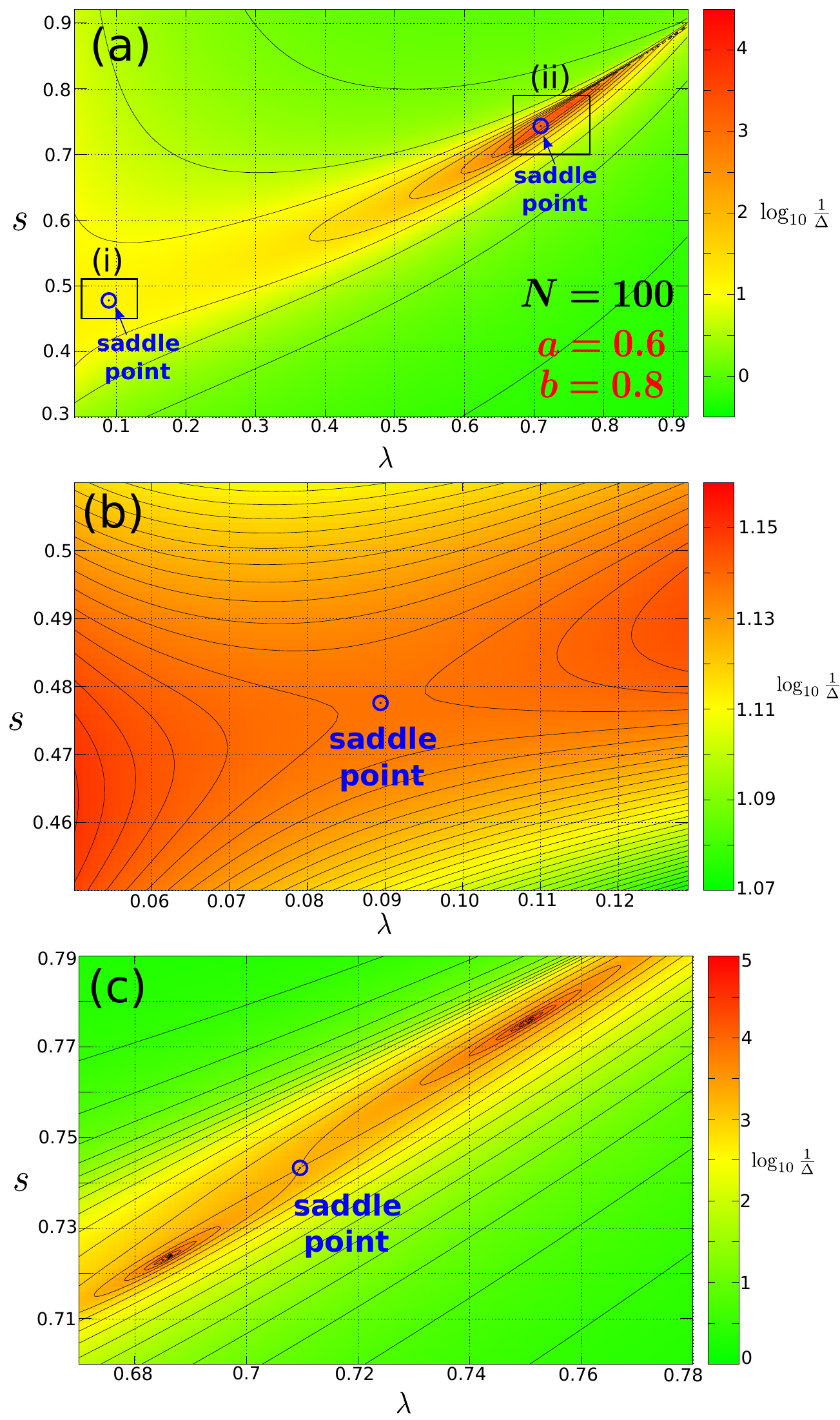}
\caption{Density plots of $\log_{10}(1/\Delta)$ of the Hamiltonian Eq. (\ref{eq.sec.III.01}) as a function of the parameters $s$ and $\lambda$, for the case of $a=0.6,\,b=0.8$ and system size $N=100$. (a) Two saddle points are identified on the gap landscape: One in Box (i), and another in Box (ii) where the gap is significantly smaller than in Box (i). Panels (b) and (c) show, respectively, close-up views of the gap landscapes in the vicinity of the saddle points of Boxes (i) and (ii).}
\label{fig.03}
\end{center}
\end{figure}

\begin{figure}[htb]
\begin{center}
\includegraphics[scale=0.7]{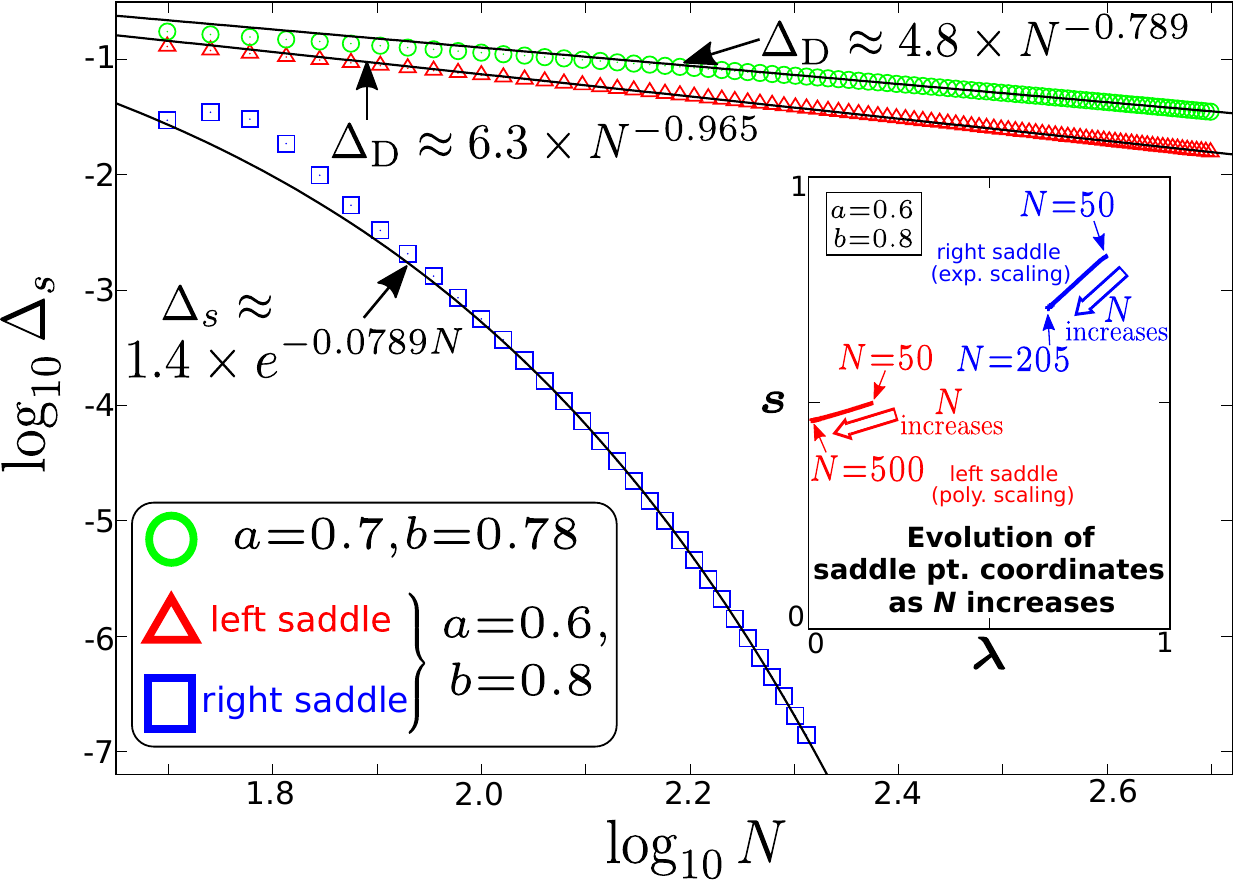}
\caption{Scalings of the energy gap for the model Eq. (\ref{eq.sec.III.01}), where the gaps, denoted $\Delta_s$, are evaluated at the saddle points on the gap landscape. If the scaling of $\Delta_s$ is polynomial, we denote it differently, as $\Delta_{\mathrm{D}}$. Results for the cases of large ($a=0.6,\,b=0.8$) and small ($a=0.7,\,b=0.78$) barrier heights are shown. The solid curve passing through the blue squares is obtained by fitting an exponential function through the data points from $N=100$ to 200. The solid lines passing through the red triangles and green circles are obtained by fitting a polynomial function through the data points from $N=100$ to 500 and from $N=200$ to 500, respectively. Inset: Evolution of the positions of the left and right saddle points on the $\lambda$-$s$ plane as $N$ increases (for $a=0.6,\,b=0.8$). As the saddle points between two successive $N$ values are quite close together, they appear to merge into a continuous line. To help the reader visualize, the starting and ending points of the lines as well as the direction of $N$ increase are indicated with arrows.}
\label{fig.04}
\end{center}
\end{figure}

The Hamiltonian $H_{\mathrm{II}}$ again commutes with the total angular momentum operator, so the numerical methods used in Sec. \ref{sec.II} are applicable to this model. Figure \ref{fig.03}(a) shows the plot of $\log_{10}(1/\Delta)$ as a function of $s$ and $\lambda$ for the parameter values $a=0.6$, $b=0.8$ with $N=100$.
An interesting observation is that, unlike the first model, we were able to identify two saddle points on the gap landscape: One inside Box (i) and another inside Box (ii). We first discuss the left one in Box (i) which has a larger gap. Figure \ref{fig.03}(b) shows a close-up view of the landscape in the vicinity of the left saddle. We tracked this point as $N$ increases; in the inset of Fig. \ref{fig.04}, the left (red) curve shows how the location of this left saddle evolves on the $\lambda$-$s$ plane as $N$ increases.
The scaling of the gap evaluated at the saddle, $\Delta_{\mathrm{D}}$, is plotted in Fig. \ref{fig.04} using red triangles; the solid black line passing through them is obtained by fitting the data points to a polynomial function. It is seen that $\Delta_{\mathrm{D}}$ decreases polynomially with increase in $N$. Hence, one can achieve a polynomial computation time for this model if the annealing path passes through this left saddle point along the line of thermodynamic first-order transitions.

\subsection{Saddle point with exponential scaling of the energy gap}
\label{sec:saddlepoint_exp}

We next discuss the right saddle point lying inside Box (ii) of Fig. \ref{fig.03}(a). Figure \ref{fig.03}(c) shows a close-up view of the gap landscape in its vicinity. This saddle has a smaller gap compared to the left one and lies between two minima on the landscape. The right (blue) curve in the inset of Fig. \ref{fig.04} shows how the location of this saddle evolves on the $\lambda$-$s$ plane as $N$ increases. The blue squares in Fig. \ref{fig.04} show the scaling of the gap evaluated at this saddle, and it closes much faster than the gap at the left saddle. Let us denote the gap here as $\Delta_s$ to differentiate it from the gap with polynomial scaling $\Delta_{\mathrm{D}}$. The solid black line passing through the squares is obtained by fitting an exponential function of the form $d_1 \exp\left(-d_2 N\right)$ ($d_1,d_2:$ fitting parameters) through the data points. It is seen that $\Delta_s$ decreases exponentially with increase in system size $N$. Hence, the computation time remains exponential if the annealing path were to pass through the right saddle point.

\subsection{Smaller energy barrier in the target Hamiltonian}
\label{sec:smaller_barrier}

The gap landscape of this model is dependent on the parameters values $a$ and $b$ chosen. To illustrate it, consider $a=0.7$, $b=0.78$, which gives a smaller energy barrier in the target Hamiltonian compared to the case of $a=0.6$, $b=0.8$. We found that although the left saddle point still exists, this time the right saddle point does not. The scaling of the energy gap at this left saddle is calculated and the result is plotted in Fig. \ref{fig.04} using green circles. The solid black line passing through the circles is obtained by fitting a polynomial function through the data points. It is seen that the gap again decreases polynomially with increase in $N$. The rate of decrease is also slower compared to the previous case of $a=0.6, b=0.8$, which is consistent since a target Hamiltonian with a smaller barrier should be easier to solve.

\section{The 3-spin ferromagnet in random longitudinal field}
\label{sec.IV}

The third model we study is the 3-spin ferromagnet in a random longitudinal field. This is a more difficult problem than the case without a random longitudinal field discussed in Ref. \cite{Durkin19} and Sec. \ref{sec.II} because the non-stoquastic driver fails to reduce first-order transitions into second order even in the $p$-spin ferromagnet with $p\ge 4$ \cite{Ichikawa:Thesis:2014}.  Remember that the reduction of the order of transition is achieved by non-stoquastic driver for the $p(\ge 4)$-spin model in the absence of a random longitudinal field \cite{Seki12,Seoane12,Nishimori17}.

The target Hamiltonian is
\begin{equation}
H_L(h_0)
=
-
{J_z}^3
-
h_0
L_z
\label{eq.sec.IV.01}
\end{equation}
where $h_0$ is the strength of the longitudinal field given by
\begin{equation}
L_z
=
\frac{1}{N}
\sum_{i=1}^N \xi_i \sigma_i^z
\label{eq.sec.IV.02}
\end{equation}
The random variables $\xi_i$ are each $+1$ or $-1$ with probability $1/2$. When $h_0=0$, we recover the 3-spin ferromagnet, which QA can solve in polynomial time using coherent catalysis. In this section, we shall discuss the case where exactly $N/2$ of the $\xi_i$ are $+1$ (we consider even $N$ in this section). Our annealing Hamiltonian is
\begin{equation}
\frac{1}{N}H_{\mathrm{III}}(s,\lambda,h_0)
=
s
\left[
\lambda
H_L(h_0)
+
(1-\lambda){J_x}^2
\right]
-(1-s)J_x,
\label{eq.sec.IV.03}
\end{equation}
where the parameters $s, \lambda$ as well as the starting and ending points of the annealing schedule are the same as for the previous model in Sec. \ref{sec.III}. According to an earlier mean-field study, the system must undergo an unavoidable first-order phase transition during the annealing process \cite{Ichikawa:Thesis:2014}.

Unlike the two previous models, the Hamiltonian $H_{\mathrm{III}}$ does not commute with the total angular momentum operator due to the longitudinal field $L_z$. To avoid diagonalizing the entire Hamiltonian matrix of dimension $2^N$, we note that $H_{\mathrm{III}}$ exhibits permutation symmetry among the spin indices with the same value of $\xi_i$. We renumber the indices $i$ such that $i=1$ to $N/2$ now have $\xi_i=+1$, and $i=N/2+1$ to $N$ now have $\xi_i=-1$, which is possible because of the symmetry of the first term $-{J_z}^3$ in the target Hamiltonian Eq. (\ref{eq.sec.IV.01}) under the permutation of spin index. The Hamiltonian can be rewritten as

\begin{eqnarray}
&&H_{\mathrm{III}}(s,\lambda,h_0)\nonumber\\[10pt]
&=&
-\frac{s\lambda}{N^2}
\left[
{\tilde{J}_z}{}^3\otimes \tilde{I}
+
3
{\tilde{J}_z}{}^2
\otimes
\tilde{J}_z
+
3
\tilde{J}_z
\otimes
{\tilde{J}_z}{}^2
+
\tilde{I}\otimes {\tilde{J}_z}{}^3
\right]
\nonumber\\[10pt]
&&
-\frac{s\lambda h_0}{2}
\left[
\tilde{J}_z\otimes \tilde{I}
-
\tilde{I}
\otimes
\tilde{J}_z
\right]
\nonumber\\[10pt]
&&
+\frac{s(1-\lambda)}{N}
\left[
{\tilde{J}_x}{}^2\otimes \tilde{I}
+
2 \tilde{J}_x\otimes \tilde{J}_x
+
\tilde{I} \otimes {\tilde{J}_x}{}^2
\right]
\nonumber\\[10pt]
&&
-(1-s)
\left[
\tilde{J}_x\otimes \tilde{I}
+
\tilde{I}
\otimes
\tilde{J}_x
\right]
\label{eq.sec.IV.04}
\end{eqnarray}
In the tensor product $A\otimes B$, the operator $A(B)$ lies in the Hilbert space of the spins whose $\xi_i$ is $+1(-1)$. $\tilde{I}$ denotes the identity operator, $\tilde{J}_{\alpha}\otimes \tilde{I}=\sum_{i=1}^{N/2} \sigma_i^{\alpha}$, and $\tilde{I}\otimes\tilde{J}_{\alpha}=\sum_{i=N/2+1}^{N} \sigma_i^{\alpha}$. The Hamiltonian $H_{\mathrm{III}}$ commutes with the total angular momentum operators of each subspace, $\sum_{\alpha} \tilde{J}_{\alpha}{}^2 \otimes \tilde{I}$ and $\tilde{I} \otimes \sum_{\alpha} \tilde{J}_{\alpha}{}^2$. To obtain just the ground and first excited states, one can diagonalize $H_{\mathrm{III}}$ using the product basis $|N/4,m_A\rangle \otimes |N/4,m_B\rangle$ where $|j,m\rangle$ is an angular momentum eigenstate with total quantum number $j$ and projection quantum number $m$. The dimension of the Hamiltonian matrix that we need to diagonalize is therefore $(N/2+1)^2$ instead of $2^N$. In our calculations, we diagonalized $H_{\mathrm{III}}$ using the Lanczos algorithm and computed only the ground and first excited states. 

\subsection{Absence of coherent catalysis when $h_0=1$}
\label{sec:h_01}

\begin{figure}[h]
\begin{center}
\includegraphics[scale=0.7]{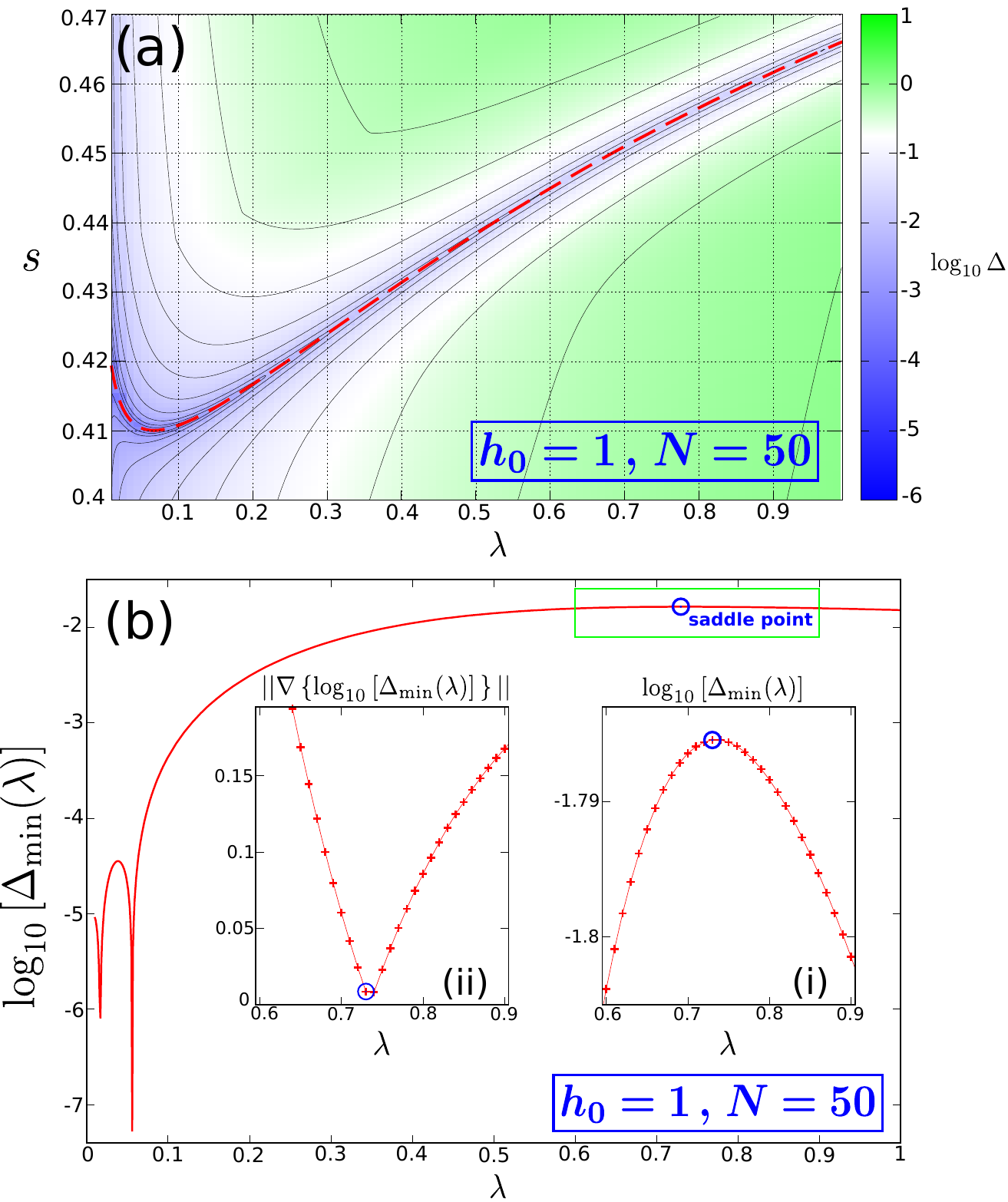}
\caption{Energy gap of the random field model Eq. (\ref{eq.sec.IV.03}) with $h_0=1$ and $N=50$. (a) Density plot of $\log_{10}\Delta$ as a function of the parameters $s$ and $\lambda$. The broken red line marks, for each $\lambda$, the $s$ where the gap is the smallest. The gap evaluated along this broken red line, denoted as $\log_{10}\left[\Delta_{\mathrm{min}}(\lambda)\right]$, is shown in Panel (b). The saddle point (blue circle) indicates where the gap is maximum along the entire curve. Inset (i): Close-up view of the gap in the vicinity of the saddle point. Inset (ii): The corresponding magnitude of the gradient of gap (see main text for details). In both insets, the connecting lines are to guide the eye only.}
\label{fig.05}
\end{center}
\end{figure}

\begin{figure}[htb]
\begin{center}
\includegraphics[scale=0.65]{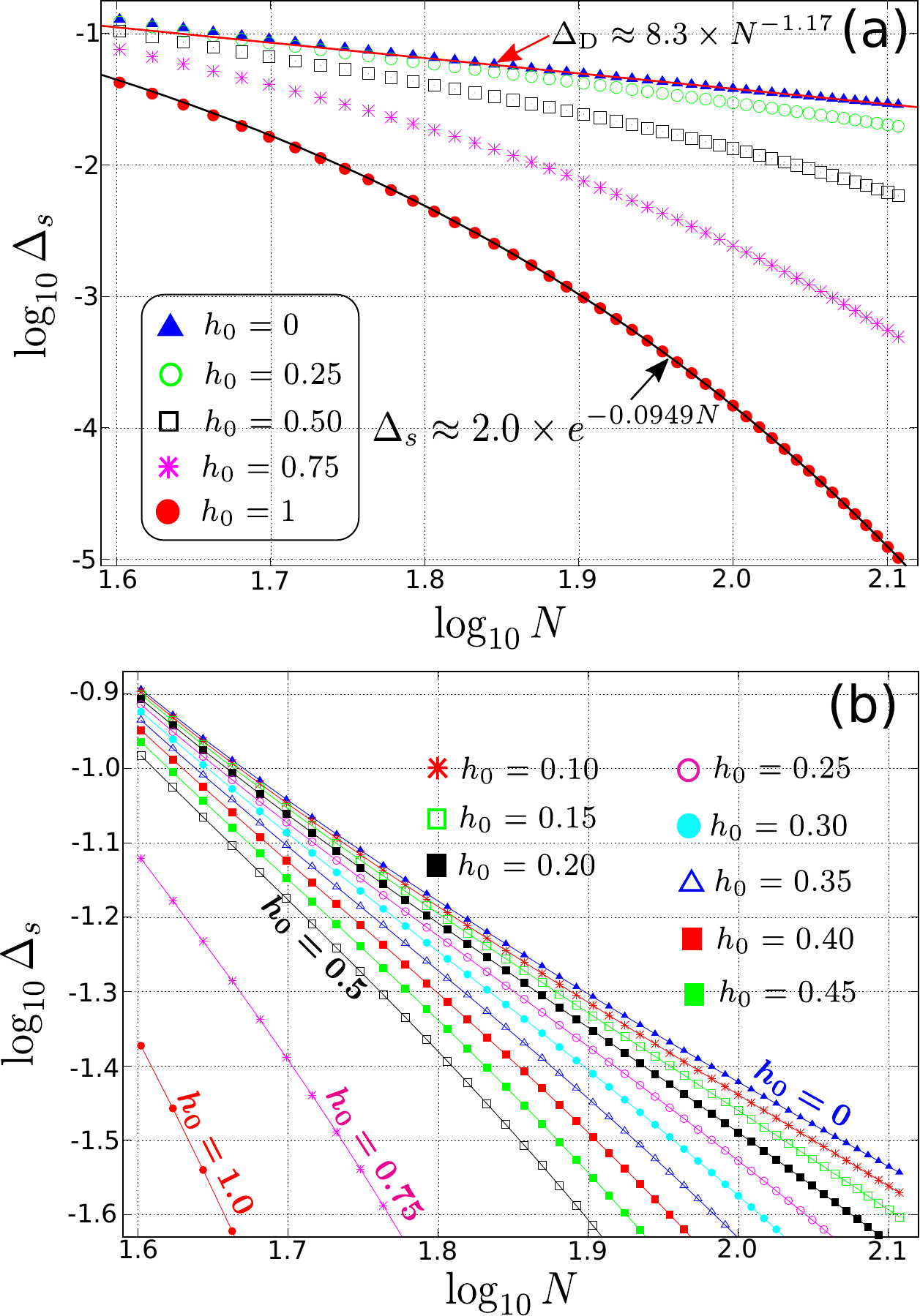}
\caption{Scaling of the energy gap for the model Eq. (\ref{eq.sec.IV.03}) for different values of $h_0$. For fixed $h_0$ and $N$, the gap $\log_{10}\Delta_s$ is evaluated at the saddle point discussed in Fig. \ref{fig.05}. (a) As $h_0$ increases from 0 to 1, the scaling of the gap changes from polynomial to exponential. The bottom (top) solid curve passing through the filled circles (triangles) is obtained by fitting an exponential (polynomial) function through the data points from $N=70$ to 128. (b) Scalings of the gap for values of $h_0$ close to $0$. There exists some non-zero values of $h_0$ ($\lesssim 0.25$) where the scaling appears to be polynomial. Connecting lines are to guide the eye only. }
\label{fig.06}
\end{center}
\end{figure}

The valley in the gap landscape in the random-field model is very sharp and numerical computations require extremely careful steps. Figure \ref{fig.05}(a) shows the density plot of the logarithm of the energy gap $\log_{10}\Delta$ as a function of $s$ and $\lambda$ for $h_0=1$ with $N=50$.
For this model, we cannot identify the saddle point on the gap landscape just by visual inspection because the minimum gap curve is sandwiched within an extremely narrow region of the $\lambda$-$s$ parameter space
\footnote{The Newton-Raphson algorithm used previously is no longer suitable for this model because (i) it is difficult to provide a good initial condition which lies in the basin of attraction of the saddle point; (ii) the change of $\log_{10}\Delta$ with respect to $\lambda$ and $s$ close to the minimum gap curve is so abrupt that the finite difference method cannot give an accurate estimate of the gradient and hessian of $\log_{10}\Delta$.}.
To locate the saddle point, we first compute the minimum gap curve
\footnote{We computed the minimum gap curve by performing a Monte Carlo search along the $s$-direction for each $\lambda$. We begin with a large step size $\delta s=2^{-2}$ and accept a proposed $s$ only if the proposed gap is smaller. We halve $\delta s$ after every $M$ ($\approx 250$) iterations until $\delta s\approx 2^{-25}$. Larger system sizes should require a larger $M$ and a smaller final $\delta s$ to arrive at a better minimum gap, but we used these $M$ and $\delta s$ values for all $N$ in our calculations. The advantage of such a Monte Carlo approach is that it requires only evaluation of the gap but not its derivatives (which diverges numerically in the vicinity of the minimum gap curve).},
which is shown in Fig. \ref{fig.05}(a) by the broken red curve. The gap evaluated along the curve, denoted as $\log_{10}\left[\Delta_{\mathrm{min}}(\lambda)\right]$, is shown in Fig. \ref{fig.05}(b). It is seen to be rich in structures such as local maxima and cusp-shaped minima. We shall focus on the global maximum indicated by a blue circle, which is also a saddle point. Inset (i) shows a close-up view of the gap in the vicinity of the maximum. To verify that the maximum is a saddle point, we took the gradient $\nabla=\left(\partial/\partial \lambda,\partial/\partial s\right)$ of the function $\log_{10}\Delta (\lambda,s)$ along the minimum gap curve
\footnote{The derivatives of the energy eigenvalues (i.e., $\partial E_0/\partial \lambda,\partial E_0/\partial s, \partial E_1/\partial \lambda$, and $\partial E_1/\partial s$) needed to compute $\nabla [ \log_{10}\Delta (\lambda,s)]$ are evaluated using the Hellmann-Feynman theorem.}.
Inset (ii) shows that the Euclidean norm of the gradient, $||\nabla ( \log_{10}\Delta ) ||$, vanishes at the maximum. The scaling of the energy gap evaluated at this saddle point, denoted as $\Delta_s$, is plotted in Fig. \ref{fig.06}(a) using solid red circles.
The solid black curve passing through the circles is obtained by fitting an exponential function through the data points. It is seen that $\Delta_s$ decreases exponentially with increase in $N$. Hence, it is not possible to reduce the gap scaling from exponential to polynomial using coherent catalysis when $h_0=1$.

\subsection{Crossover from exponential to polynomial scaling between $h_0=1$ and $0$ }
\label{sec:crossover}

As mentioned earlier, when $h_0=0$ the target Hamiltonian Eq. (\ref{eq.sec.IV.01}) reduces to the 3-spin ferromagnet whose energy gap at the saddle point exhibits polynomial scaling. This is shown in Fig. \ref{fig.06}(a) using solid blue triangles; the solid red line passing through the triangles is obtained by fitting a polynomial function through the data points. The scaling of the saddle point gap for $h_0=0.25$, 0.50, and 0.75 is also shown in Fig. \ref{fig.06}(a). Figure \ref{fig.06}(b) shows a detailed view of the scaling curves, with values of $h_0$ close to 0. The gap scaling seems to change from polynomial to exponential as $h_0$ is increased from 0 to 1. 

\begin{figure}[htb]
\begin{center}
\includegraphics[scale=0.65]{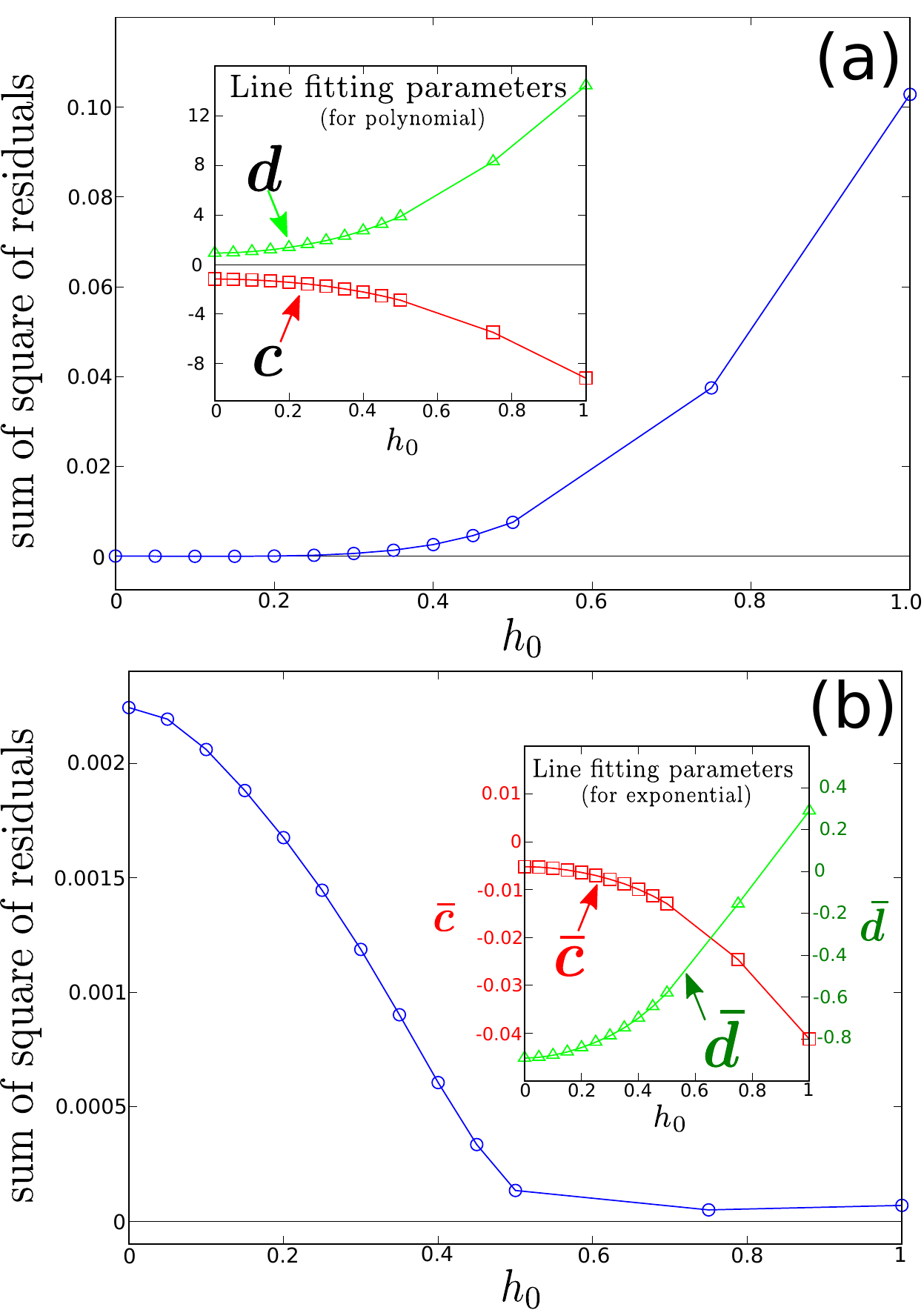}
\caption{Crossover of the scaling of the gap from polynomial to exponential as $h_0$ increases from 0 to 1. All connecting lines are to guide the eye only. (a) Fitting the gap scaling of each $h_0$ to the polynomial function Eq. (\ref{eq.sec.IV.05}). The goodness of fit is given by the sum of the square of the residuals. For $h_0\lesssim 0.25$, the data points obey the polynomial scaling quite well. The fitting parameters $c$ and $d$ obtained are shown in the inset. (b) Fitting the gap scaling of each $h_0$ to the exponential function Eq. (\ref{eq.sec.IV.06}). Although the exponential fit is good for $h_0\gtrsim 0.5$, the goodness of fit decreases for smaller values of $h_0$, especially close to 0. The fitting parameters $\bar{c}$ and $\bar{d}$ obtained are shown in the inset.}
\label{fig.07}
\end{center}
\end{figure}

To quantify this observation, let us fit the scaling data of the gap $\Delta_s$ of each $h_0$ to a polynomial curve of the form
\begin{equation}
\log_{10}\Delta_s=c\log_{10}N + d
\label{eq.sec.IV.05}
\end{equation}
The fitting parameters $c$ and $d$ obtained are shown in the inset of Fig. \ref{fig.07}(a).
How well the data points are fitted by the polynomial function Eq. (\ref{eq.sec.IV.05}) is given by the sum of the square of residuals
\footnote{The residual of the $i$th data point $d_i$ is defined as $r_i=d_i-f$ where $f$ is the fitted function. The sum of the square of residuals is $\sum_i (r_i)^2$. In our calculations, the curve fitting of the scaling data of each $h_0$ was performed using data points from $N=70$ to 128.},
which is shown in Fig. \ref{fig.07}. It is seen that for $h_0\lesssim 0.25$ the polynomial function fits the data very well; as $h_0$ increases beyond 0.25 the residual increases rapidly, indicating a deterioration of the fit. For comparison, we repeated the curve fitting, this time to an exponential function of the form
\begin{equation}
\log_{10}\Delta_s= \bar{c}\, N + \bar{d}
\label{eq.sec.IV.06}
\end{equation}
The fitting parameters $\bar{c}$ and $\bar{d}$ obtained are shown in the inset of Fig. \ref{fig.07}(b), and the sum of the square of residuals are shown in Fig. \ref{fig.07}. It is seen that although for $h_0\gtrsim 0.5$ the exponential form fits the data well, for $h_0 < 0.5$ the residual increases rapidly, indicating a deterioration of the fit for small values of $h_0$.

To conclude, we found evidence that there exist a range of values of $h_0$ where the scaling of the saddle point gap is likely to be polynomial.  Coherent catalysis works to reduce computational complexity in this difficult problem.

\section{Summary and discussions}
\label{sec.V}

We have studied the possibility of avoiding an exponentially closing energy gap implied by first-order phase transitions in quantum annealing using the mechanism of coherent catalysis recently proposed by Durkin \cite{Durkin19}. The models studied here are difficult for QA in the sense that even with non-stoquastic drivers, the mean-field theory predicts that a first-order phase transition is unavoidable during the annealing process. Coherent catalysis, on the other hand, predicts the existence of a saddle point on the energy gap landscape where the exponential gap scaling is softened into polynomial. Our purpose is to investigate, via numerical calculations, whether this phenomenon is not specific to the 3-spin ferromagnet but also occurs in other systems as well. In all the three models studied, we have found the effectiveness of coherent catalysis, i.e. the existence of a saddle point in the phase diagram where the gap scales polynomially though the thermodynamic phase transition is of first order. It was argued in Ref. \cite{Durkin19} that coherent catalysis represents intrinsic quantum effects in the mean-field-type models, and thus the present results may be understood as additional examples of quantum enhancement in optimization algorithms.

In this work, we have approached coherent catalysis purely from an empirical (i.e. numerical) perspective. Some theoretical aspects of the mechanism, however, remain unclear. For instance, we have seen from the models in Secs. \ref{sec.III} and \ref{sec.IV} that a saddle point on the gap landscape can exhibit polynomial or exponential gap scaling, not just the former. Hence, the existence of such a point does not automatically imply the reduction of the scaling, and it would be interesting to understand the conditions under which polynomial scaling will occur. Another issue concerns the type of models that has been investigated. Both here and in Ref. \cite{Durkin19}, coherent catalysis was demonstrated mainly using mean-field type models. An interesting question would be if it can occur in finite-dimensional or spin-glass systems as well. We hope to address some of these questions in our future efforts.

\begin{acknowledgements}
We thank Gabriel Durkin and Yu Yamashiro for stimulating discussions. This work is based on a project commissioned by the New Energy and Industrial Technology Development Organization (NEDO).
\end{acknowledgements}

\bibliography{references.bib}

\end{document}